\documentstyle[12pt,epsfig]{article}
\begin{document}
\title{Decoherence and time emergence.}

\author{ A. Camacho 
\thanks{email: abel@abaco.uam.mx}\\
Physics Department, \\
Universidad Aut\'onoma Metropolitana-Iztapalapa. \\
P. O. Box 55-534, 09340, M\'exico, D.F., M\'exico.}

\date{}
\maketitle

\begin{abstract} In this work the possible role that Decoherence Model could play in the emergence of the classical concept of 
time is analyzed. 
 We take the case of a Mixmaster universe with small anisotropy and construct its Halliwell propagator. Afterwards we introduce in our 
system terms that comprise the effects of Decoherence Model. This is done by means of the so called Restricted Path Integral Formalism. We obtain Halliwell's modified propagator and find that a gauge invariant physical time emerges as consequence of this process. 
\end{abstract}
\bigskip
\bigskip

\section {Introduction}

Decoherence Model [1] (DM) pretends to solve some of the old conundrums that, since its inception, beset Quantum Theory (QT). For instance, 
the measurement problem, 
namely, the appearance of a classical bahaviour starting from the superposition of several, \-ma\-cros\-co\-pic\-ally\- different, quantum states. 

Let us now try to analyze the possible role that DM could play in the solution of some very important conceptual difficulties 
that we now face in the search of a Quantum Theory of Gravity. If we take a look at the \-cu\-rrent\- efforts that in this direction we already have [2], we may immediately notice that in all of them there is always 
present a very strong restriction, they all found their begining assumptions taking as valid the usual QT. 
This seems not only very reasonable, but also soundly justified. But we should not then forget, that if we could obtain a Quantum Theory 
of Gravity, along these ideas, then they will inherit all the conceptual problems that plague QT. In particular, we may have the possible hypothetical 
situation, it is possible to have the superposition of two quantum states associated to two, macroscopically different, configurations of the gravitational field. 
In other words, we would 
face the old \-pro\-blem\- of Schr\"odinger's cat. Clearly, we do employ every day QT and do not worry very much about this problem, 
maybe because sometimes we find von Neumann postulate good enough [3]. But in the 
case of a possible quantization of the gravitational field, the situation could be worse than usual. 
The reason stems from the fact that it has already been proved that von Neumann postulate is incompatible with Special Relativity [4]. 
Thus, we would have, on one hand, this supposed Quantum Theory of Gravity, and on the other one, von Neumann postulate, 
and therefore, we would thus, unavoidably,  run into conceptual troubles.

As mention before, DM pretends to explain the appearance of classical properties starting from QT. Therefore, we may, at this point, wonder 
if it could give a possible solution to some of the problems that in this direction we face in the quest of a Quantum Theory of 
Gravity, for instance, the time problem, but also at the same time 
employ a quantization scheme that, at least in principle, could avoid the aforementioned problems of the usual efforts.

Let us be a little more explicit.  The classical diffeomorphism invariance of general relativity leads to the presence of constraints: the total Hamiltonian must vanish, then the wave functional obeys the Wheeler--De Witt equation, $H\psi = 0$. 
Since due to the uncertainty relations no spacetimes exist anymore at the level of quantum gravity, there is no time parameter available, with other words, the Wheeler--DeWitt equation is timeless, one may say that there is no time in quantum cosmology. 

In this work we will introduce DM in a particular quantization scheme (Halliwell's path--integral quantization [5]) and analyze its possibilities in connection with this time problem. 
The introduction of DM could also give us the possibility to explain the gravitational version of Schr\"odinger's cat problem.
The essential point in the application of DM on the case of quantum cosmology is that gravity couples to all forms of energy, gravity is measured by matter and therefore a general superposition of gravitational quantum states is decohered [6].  
In other words, if we consider the continuous measurement of the quantum universe, then its dynamics may be modified in such a way that time arises. 
The role of measuring device is played by higher multipoles of matter [6], which describe density fluctuations and gravitational waves present in the universe. 
These higher multipoles may thereby be considered as the environment associated to the superspace variables of the model, which in this proposal play the role of collective variables.

The mathematical formalism used to consider DM is the so called Restricted Path--Integral Formalism (RPIF) [7]. 

We will show that time emerges as a quantitative feature of our model, namely a gauge invariant physical time emerges as 
consequence of the self-measurement process of the universe. 
\bigskip

\section {Propagation amplitudes.}

Let us consider the Mixmaster metric

\begin{equation}
 {ds^2 = -N^2d^2\tau + e^{2\alpha}(e^{2\beta})_{ij}\sigma^i\sigma^j  }, 
\end{equation} 
\noindent where $\beta_{ij}$ are the elements of a traceless diagonal matrix, $N, \alpha$ and $\beta_{ij}$ are functions only of $\tau$, with $\sigma^1 = cos\varphi d\theta + sin\varphi sin\theta d\phi$, $\sigma^2 = sin\varphi d\theta - cos\varphi sin\theta d\phi$, $\sigma^3 = d\varphi + cos\theta d\phi$ and $det(e^{2\beta}) = 1$. We have the geometry of a homogeneous but not isotropic sphere [8].

Here $\tau$ is an arbitrary parameter related to the foliation of the classical spacetime into spatial hypersurfaces, and if it suffers the action of a transformation, namely if we have $d\tau \rightarrow d\zeta = f(\tau)d\tau$, then invariance demands also the transformation of the lapse function, we must also carry out the transformation $N(\tau) \rightarrow M(\zeta) = {N(\tau)\over f(\tau)}$. 

The (3 + 1) decomposition of the metric is [8] $g_{ij} = e^{2\alpha}e^{2\beta_{ij}}\delta_{ij}$, $N_i = 0$, $N^{\bot} = N^{-1}$ and $\pi^{ij} = {e^{\alpha - 2\beta_{ij}}\over N}(\dot{\beta}_{ij} -2\dot{\alpha)}\delta_{ij}$.

The action is 

\begin{equation}
{ S = \int (\pi^{ij}\dot{g}_{ij} - N^{\mu}H_{\mu})d^4x }, 
\end{equation} 

\noindent we use $\hbar = 1$, $c = 1$ and $G = 1$.

For this particular case we have 

{\setlength\arraycolsep{2pt}\begin{eqnarray}
 \pi ^{ij}\dot{g}_{ij} - N^{\mu}H_{\mu} =  {2\over N}e ^{3\alpha}\{(\dot{\beta}_{11} - 2 \dot{\alpha})(2\dot{\beta}_{11} + \dot{\alpha}) + (\dot{\beta}_{22} - 2 \dot{\alpha})(2\dot{\beta}_{22} + \nonumber\\
\dot{\alpha}) + (\dot{\beta}_{33} - 2 \dot{\alpha})(2\dot{\beta}_{33} 
+ \dot{\alpha}) - {1\over 2}[3\dot{\beta}^2_{33} + (\dot{\beta}_{11} - \dot{\beta}_{22})^2 -12\dot{\alpha}^2]\} + \nonumber\\
{N\over 2}e ^{2\alpha}Tr(2e ^{-2\beta} -e ^{4\beta}).
\end{eqnarray}

The evolution of a quantum model of the universe may be described \'a la Halliwell [5] by the following propagator 

\begin{equation}
{ U(q'', q') = (\tau '' - \tau ')\int d[N^{\bot}]\delta(\dot{N}^{\bot})d[p_l]d[q^l]exp\{i\int d\tau [p_l\dot{q}^l - N^{\bot}H_{\bot}]\}  }, 
\end{equation} 

\noindent where $q'' = (\alpha'', \beta'', \tau'')$ and $q' = (\alpha', \beta', \tau')$.

From this last expression we may evaluate the probability transition $P = \vert U \vert^2$ from the initial configuration $q'$ to the final $q''$.

 For our metric we may write the propagator as 

{\setlength\arraycolsep{2pt}\begin{eqnarray}
 U(q'', q') = (\tau '' - \tau ')\int dNd[p_+]d[p_-]d[p_{\alpha}]d[\alpha]d[\beta_+]d[\beta_-]\nonumber\\
exp\Bigl[i\int_ {\tau '}^{\tau ''}\{p_+\dot{\beta}_+ + p_-\dot{\beta}_- \nonumber\\
- p_{\alpha}\dot{\alpha} - {N\over 3\pi}e ^{-3\alpha}(p^2_+ + p^2_- - p^2_{\alpha}) + \nonumber\\
{3\pi N\over 2}e ^{\alpha - 2\beta_+}(e ^{-\sqrt{12}\beta_-} + e ^{\sqrt{12}\beta_-} 
+ e ^{6\beta_+}) - \nonumber\\
{3\pi N\over 4}e ^{\alpha + 4\beta_+}(e ^{-4\sqrt{3}\beta_-} + 
e ^{4\sqrt{3}\beta_-} + e ^{-12\beta_+})\}d\tau\Bigr],
 \end{eqnarray} 
\noindent where $\beta_- = {1\over 2\sqrt{3}}(\beta_{11} - \beta_{22})$, $\beta_+ = {1\over 2}(\beta_{11} + \beta_{22})$ and $\beta_{33} = -2\beta_+$ [8].

First the integrals with respect to the momenta will be carried out, and in this integration we use the result $\int d[p]exp\{- {1\over 2}([p], A[p]) + ([q], [p])\} = exp\{{1\over 2}([q], A^{-1}[q]) \}$ [7], where $([q], [p]) = \int_ {\tau '}^{\tau ''}q(\tau)p(\tau)d\tau$.

{\setlength\arraycolsep{2pt}\begin{eqnarray}
\int d[p_+]d[p_-]d[p_{\alpha}]exp\Bigl[i\int_ {\tau '}^{\tau ''} d\tau \{p_+\dot{\beta}_+ + p_-\dot{\beta}_-  - p_{\alpha}\dot{\alpha}  \nonumber\\
+ {N\over 3\pi}e ^{-3\alpha}(p^2_{\alpha} - p^2_+ - p^2_-)\}\Bigr] =   exp\Bigl[{3i\pi\over 4N}\int_ {\tau '}^{\tau ''}\{\dot{\beta}^2_+ + \dot{\beta}^2_- - \dot{\alpha}^2\}e ^{3\alpha}d\tau\Bigr] .
\end{eqnarray} 

Then the propagator becomes 

{\setlength\arraycolsep{2pt}\begin{eqnarray}
 U(q'', q') = (\tau '' - \tau ')\int dNd[\alpha]d[\beta_+]d[\beta_-]exp\Bigl[i\int_ {\tau '}^{\tau ''}\{{3\pi N\over 2}e ^{\alpha-2\beta_+}\cr
\times (e ^{-\sqrt{12}\beta_-} +  e^{\sqrt{12}\beta_-} +  e ^{6\beta_+}) -  {3\pi N\over 4}e ^{\alpha + 4\beta_+}(e ^{-4\sqrt{3}\beta_-} + e ^{4\sqrt{3}\beta_-} + e ^{-12\beta_+}) \nonumber\\ 
+ {3\pi\over 4N}e^{3\alpha}(\dot{\beta}^2_+ + \dot{\beta}^2_- - \dot{\alpha}^2)\}\Bigr]d\tau  .
\end{eqnarray} 

In order to obtain an analytical expression for our propagator let us now consider a more symmetric case, we will introduce two restrictions $\beta_- = 0$ and $0 < \vert\beta_+ \vert \ll 1$. 
Hence, the resulting Hamiltonian acquires a very simple form [9] and in consequence our propagator becomes (from now on we drop the subindex of $\beta_+$)

\setlength\arraycolsep{2pt}\begin{eqnarray}
 U(q'', q') \cong (\tau '' - \tau ')\int dNd[\alpha]d[\beta]exp\Bigl[i\int_ {\tau '}^{\tau ''}\{{3\pi N\over 4}e ^{\alpha}(1 - 8\beta^2) \nonumber\\
+ {3\pi\over 4N}e^{3\alpha}(\dot{\beta}^2 - \dot{\alpha}^2)\}d\tau\Bigr] .
\end{eqnarray}

As an additional approximation we will take only terms up to second order in $\alpha$, namely $e^{\alpha} \cong 1 + \alpha + {\alpha^2\over 2}$. 
Considering this approximation we may now rewrite the expression for the propagator as follows

{\setlength\arraycolsep{2pt}\begin{eqnarray}
U(q'', q') = (\tau '' - \tau ')\int dNd[\alpha]d[\beta]exp\Bigl[\int_ {\tau '}^{\tau ''}\{{3i\pi N\over 4}(1 + \alpha + {\alpha^2\over 2}) \nonumber\\
- 6\pi iN\beta^2 + {3i\pi \over 4N}(\dot{\beta}^2 - \dot{\alpha}^2)\}d\tau\Bigr]. 
\end{eqnarray} 

Let us consider first the integral

{\setlength\arraycolsep{2pt}\begin{eqnarray}
\int d[\alpha]exp\Bigl[\int_ {\tau '}^{\tau ''}\{{3i\pi N\over 4}(1 + \alpha + {\alpha^2\over 2}) - {3i\pi \over 4N}\dot{\alpha}^2\}d\tau\Bigr] = exp\{{3i\pi N\over 4}(\tau '' - \tau ')\}\nonumber\\
\times \int d[\alpha] exp\Bigl[i\int_ {\tau '}^{\tau ''}\{{1\over 2}(-{3\pi \over 2N})\dot{\alpha}^2 - {1\over 2}(-{3\pi \over 2N})({N^2\over 2}){\alpha}^2 + {3\pi N\over 4}\alpha \}d\tau\Bigr].
\end{eqnarray} 

The functional integral on the right-hand side of (10) may be understood as the propagator of a driven harmonic oscillator, with mass $m = -{3\pi \over 2N}$, frequency $\omega = {N\over\sqrt{2}}$ and where the external force is $F(\tau) = {3\pi N\over 4}$.

The ``classical action'' of this system is [10]

{\setlength\arraycolsep{2pt}\begin{eqnarray}
S_{\alpha} = {-3\pi\over 4\sqrt{2}sin\Bigl({N\over\sqrt{2}}(\tau '' - \tau ')\Bigr)}\Bigl[(\alpha''^2  + \alpha'^2 )cos\Bigl({N\over\sqrt{2}}(\tau '' - \tau ')\Bigr) \nonumber\\
- 2\alpha''\alpha' -4(\alpha'' + \alpha')sin^2\Bigl({N\over\sqrt{8}}(\tau '' - \tau ')\Bigr) - 4sin^2\Bigl({N\over\sqrt{8}}(\tau '' - \tau ')\Bigr) \cr 
+ {N\over\sqrt{2}}(\tau '' - \tau ')\Bigr]. 
\end{eqnarray} 

Therefore 

{\setlength\arraycolsep{2pt}\begin{eqnarray}\int d[\alpha] exp\Bigl[\int_ {\tau '}^{\tau ''}\{{3i\pi N\over 4}(1 + \alpha + {\alpha^2\over 2}) - {3i\pi \over 4N}\dot{\alpha}^2\}d\tau\Bigr] = \nonumber\\ 
\sqrt{{3i\over 4\sqrt{2}sin\Bigl({N\over\sqrt{2}}(\tau '' - \tau ')\Bigr)}}exp\{{3i\pi N\over 4}(\tau '' - \tau ') + iS_{\alpha}\}.
\end{eqnarray} 

In a similar way we have that

\begin{equation} { \int d[\beta]exp\Bigl[\int_ {\tau '}^{\tau ''}\{- 6\pi iN\beta^2 + {3i\pi \over 4N}\dot{\beta}^2\}d\tau\Bigr] = \sqrt{{3\over \sqrt{2}isin\Bigl(\sqrt{8}N(\tau '' - \tau ')\Bigr)}} exp\{iS_{\beta}\} }, \end{equation} 

\noindent where $S_{\beta}$ is the classical action of a free harmonic oscillator with mass $m = {3\pi \over 2N}$ and frequency $\omega = \sqrt{8}N$.

\begin{equation}
{ S_{\beta} = {3\pi\over \sqrt{2}sin\Bigl(\sqrt{8}N(\tau '' - \tau ')\Bigr)}\Bigl[(\beta''^2 + \beta'^2 )cos\Bigl(\sqrt{8}N(\tau '' - \tau ')\Bigr) - 2\beta''\beta'\Bigr] }. 
\end{equation}
 
From the last integrations we obtain the propagator of a quantum mixmaster universe with small anisotropy, here self-measurement has not been taken into account. 

\begin{equation} 
{  U(q'', q') = \sqrt{{9\over 8}}(\tau '' - \tau ')\int dN {exp\Bigl[ i({3\pi N(\tau '' - \tau ')\over 4} + S_{\alpha} + S_{\beta})\Bigr]\over \sqrt{sin\Bigl({N\over\sqrt{2}}(\tau '' - \tau ')\Bigr)sin\Bigl(\sqrt{8}N(\tau '' - \tau ')\Bigr)}} }. 
\end{equation} 

We now proceed to introduce self-measurement in our universe. This will be done employing RPIF which is a phenomenological approach. This last fact means that we will introduce some parameters that can not be explained in our model but the approach has the advantage that it allows us to consider the influence of the measuring device and at the same time it also enables us to forget the actual scheme of measurement. 

Self-measurement means that some functions $[\kappa]$, $[\nu]$ and $[\gamma]$ are found as estimates of the corresponding functions $[N]$, $[\beta]$ and $[\alpha]$. 

Invariance under reparametrization $d\tau \rightarrow d\zeta = f(\tau)d\tau$ implies that the weight functionals to be introduced in the path integrals must be invariant under this reparametrization. 

This invariance condition is fulfilled if we consider the following weight functionals 

\begin{equation} { \omega_{[\kappa]} = exp\{ -\int_ {\tau '}^{\tau ''}{\vert N - \kappa \vert \over \sigma^2}d\tau\}   }, \end{equation} 

\begin{equation} { \omega_{[\nu]} = exp\{ -\int_ {\tau '}^{\tau ''}{N(\nu - \beta)^2 \over \rho^2}d\tau\}   } ,\end{equation} 

\begin{equation} { \omega_{[\gamma]} = exp\{ -\int_ {\tau '}^{\tau ''}{N(\gamma - \alpha)^2 \over \Omega^2}d\tau\}   } .\end{equation} 

Clearly, we do not know if the self-measurement process of the universe renders these functionals.
 But for a qualitative analysis of the consequences of this self-measurement process in the dynamics of the universe we may neglect in a first approach the details in the definition of the involved functionals and therefore we may choose the most convenient functionals. These Gaussian weights lead to Gaussian integrals which can be easily performed.
 
A more precise treatment of this issue demands the analysis of the role that higher multipoles of matter play in the definition of the environment associated with the superspace. From this analysis we could also comprehend how the constants $\rho^2 $, $\Omega^2$ or $\sigma^2$, which in this phenomenological approach can not be explained, are defined by the density fluctuations and gravitational waves present in our universe.

Under this choice expression (9) becomes now 

{\setlength\arraycolsep{2pt}\begin{eqnarray} U_{[\kappa, \nu, \gamma]}(q'', q') = (\tau '' - \tau ')\int dNd[\alpha]d[\beta ]exp\Bigl[\int_ {\tau '}^{\tau ''}\{{3i\pi N\over 4}(1 + \alpha + {\alpha^2\over 2}) \nonumber\\
- 6\pi iN\beta^2 + {3i\pi \over 4N}(\dot{\beta}^2 - \dot{\alpha}^2) - {\vert N - \kappa \vert \over \sigma^2} - {N(\nu - \beta)^2 \over \rho^2} - {N(\gamma - \alpha)^2 \over \Omega^2}\}d\tau \Bigr].
 \end{eqnarray} 

Consider now the expression 

\begin{equation}
{ \int d[\beta ]exp\Bigl[\int_ {\tau '}^{\tau ''}\{{3i\pi \over 4N}\dot{\beta}^2 - 6\pi iN\beta^2 - {N(\nu - \beta)^2 \over \rho^2}\}d\tau\Bigr]. }
\end{equation}

It may be seen as the propagator of a free harmonic oscillator with mass $m = {3\pi \over 2N}$ and frequency $\omega = \sqrt{8}N$ under continuous measurement of its position $\beta$, such that the function $\nu(\tau)$ is obtained as result of this measurement and the error done in the position measuring is $\Delta\nu = \sqrt{{2\over \vert N(\tau '' - \tau ')\vert}}\rho$.

The propagator of this oscillator is [7, 10]

{\setlength\arraycolsep{2pt}\begin{eqnarray}
 \int d[\beta ]exp\Bigl[\int_ {\tau '}^{\tau ''}\{{3i\pi \over 4N}\dot{\beta}^2 - 6\pi iN\beta^2 - {N(\nu - \beta)^2 \over \rho^2}\}d\tau\Bigr] = \nonumber\\
\sqrt{{3\sqrt{1 - {i\over 6\pi\rho^2}}\over \sqrt{2}isin\Bigl(\sqrt{8}N\sqrt{1 - {i\over 6\pi\rho^2}}(\tau '' - \tau ')\Bigr)}}exp \Bigl[-{\vert N(\tau '' - \tau ')\vert\over \rho^2}<\nu^2> +  iS_{\beta}]\Bigr]. 
\end{eqnarray} 

Here

{\setlength\arraycolsep{2pt}\begin{eqnarray} 
S_{\beta} \cong {3\pi\Gamma\over \sqrt{2}sin\Bigl(\sqrt{8}N\Gamma (\tau '' - \tau')\Bigr)}\times\nonumber\\
\Bigl[ (\beta''^2 + \beta'^2 )cos\Bigl(\sqrt{8}N\Gamma (\tau '' - \tau')\Bigr) - \nonumber\\
2 \beta''\beta' - i{\sqrt{8}N(\tau '' - \tau')(\beta''+ \beta') \over 3\pi\rho^2\Gamma}\times\nonumber\\
\nu({\tau'' + \tau'\over 2})
sin\Bigl(\sqrt{8}N\Gamma(\tau '' - \tau')\Bigr) + \nonumber\\ 
{2N^2(\tau '' - \tau')^2\over 9\pi^2\rho^4\Gamma^2}\nu ({\tau'' + \tau'\over 2})\nu({\tau'' + 3\tau'\over 4})
\times\nonumber\\
sin\Bigl(\sqrt{2}N\Gamma(\tau '' - \tau')\Bigr)sin\Bigl({N\over\sqrt{2}}\Gamma(\tau '' - \tau')\Bigr)\Bigr],
\end{eqnarray} 

\noindent $\Gamma = \sqrt{1 - {i\over 6\pi\rho^2}}$, and $<\nu^2> = {1 \over\tau '' - \tau'}\int_ {\tau '}^{\tau ''}\nu(\tau)^2d\tau$ and where we may understand $S_{\beta}$ as the ``classical action'' of a fictitious complex driven oscillator whose mass and frequency are $m = {3\pi \over 2N}$, $\upsilon = \sqrt{8}N\Gamma$, respectively, and where the external force is $F(\tau) = -i{2N\over \rho^2}\nu(\tau)$.

In the case of the integral$\int d[\alpha]exp\Bigl[\int_ {\tau '}^{\tau ''}\{{3i\pi N\over 4}(\alpha + {\alpha^2\over 2})  - {3i\pi \over 4N}\dot{\alpha}^2- {N(\gamma - \alpha)^2 \over \Omega^2}\}d\tau\Bigr] $ the situation resembles the case of expression (21). 
Indeed, we have a harmonic oscillator with mass $m = -{3\pi \over 2N}$, frequency $\omega = {N\over\sqrt{2}}$ and under the influence of the force $F(\tau) = {3\pi N\over 4}$. 
Here the position $\alpha$ is continuously measured, and $\gamma(\tau)$ and $\Delta\gamma = \sqrt{{2\over \vert N(\tau '' - \tau ')\vert}}\Omega$ are the result and involved error in this measurement process, respectively.

The propagator for this harmonic oscillator is also easily calculated 

{\setlength\arraycolsep{2pt}\begin{eqnarray} 
\int d[\alpha]exp\Bigl[\int_ {\tau '}^{\tau ''}\{{3i\pi N\over 4}(1 + \alpha + {\alpha^2\over 2}) - {3i\pi \over 4N}\dot{\alpha}^2 -{N(\gamma - \alpha)^2 \over \Omega^2}\}d\tau \Bigr] = \cr
\sqrt{{3i\sqrt{1 + {i8\over 3\pi\Omega^2}}\over 4\sqrt{2}sin\Bigl({N\over \sqrt{2}}\sqrt{1 + {i8\over 3\pi\Omega^2}}(\tau '' - \tau ')\Bigr)}}exp\Bigl[{3i\pi N\over 4}(\tau '' - \tau ') \nonumber\\
- \vert N(\tau '' - \tau ')\vert{<\gamma^2>\over \Omega^2} + iS_{\alpha}\Bigr] . 
\end{eqnarray} 

Here we have

{\setlength\arraycolsep{2pt}\begin{eqnarray} 
S_{\alpha} \cong {-3\pi\tilde\omega\over 4\sqrt{2}sin\Bigl({N\over \sqrt{2}}\tilde\omega(\tau '' - \tau ')\Bigr)}\Bigl[(\alpha''^2 + \alpha'^2)cos\Bigl({N\over \sqrt{2}}\tilde\omega(\tau '' - \tau ')\Bigr) \nonumber\\
-2\alpha''\alpha' - 4{(\alpha'' + \alpha')\over\tilde\omega^2}sin^2\Bigl({N\over \sqrt{8}}\tilde\omega(\tau '' - \tau ')\Bigr) \nonumber\\
+ i{8\sqrt{2}N(\tau '' - \tau ')(\alpha'' + \alpha')\over 3\pi\Omega^2\tilde\omega}\gamma({\tau '' + \tau '\over 2})sin\Bigl({N\over \sqrt{8}}\tilde\omega(\tau '' - \tau ')\Bigr)\nonumber\\
- 4\tilde\omega^{-4}sin^2\Bigl({N\over \sqrt{2}}\tilde\omega(\tau '' - \tau ')\Bigr) + {N\over \sqrt{2}}{(\tau '' - \tau ')\over \tilde\omega^3}\nonumber\\
+ i{4N^2(\tau '' - \tau ')^2\over 3\pi\Omega^2\tilde\omega^2}[\gamma({\tau '' + \tau '\over 2}) + \gamma({\tau '' + 3\tau '\over 4})]\nonumber\\
\times sin\Bigl({N\over \sqrt{8}}\tilde\omega(\tau '' - \tau ')\Bigr)sin\Bigl({N\over \sqrt{32}}\tilde\omega(\tau '' - \tau ')\Bigr) \nonumber\\
+ {32N^2(\tau '' - \tau ')^2\over 9\pi^2\Omega^4\tilde\omega^2}\gamma({\tau '' + \tau '\over 2})\gamma({\tau '' + 3\tau '\over 4})\nonumber\\
\times sin\Bigl({N\over \sqrt{8}}\tilde\omega(\tau '' - \tau ')\Bigr)sin\Bigl({N\over \sqrt{32}}\tilde\omega(\tau '' - \tau ')\Bigr)\Bigr], 
\end{eqnarray}

\noindent where $\tilde\omega = \sqrt{1 + {i8\over 3\pi\Omega^2}}$ and $<\gamma^2> = {1 \over\tau '' - \tau'}\int_ {\tau '}^{\tau ''}\gamma(\tau)^2d\tau$ and here we may understand $S_{\alpha}$ as the ``classical action'' of a fictitious complex driven oscillator whose mass and frequency are $m = -{3\pi \over 2N}$, $\upsilon = {N\over\sqrt{2}}\tilde\omega$, respectively, and where the involved external force is $F(\tau) = {3\pi N\over 4} -i{2N\over \Omega^2}\gamma(\tau)$.

Therefore, the propagator with self-measurement is

{\setlength\arraycolsep{2pt}\begin{eqnarray} 
U_{[\kappa, \nu, \gamma]}(q'', q') = \sqrt{{9\over 8}}\Bigl[(1 + {i8\over 3\pi\Omega^2})(1 - {i\over 6\pi\rho^2})\Bigr]^{1\over 4}(\tau '' - \tau ')\nonumber\\
\int dN{exp\Bigl[iS + N(\tau '' - \tau '){3i\pi \over 4} -\vert N(\tau '' - \tau ')\vert({<\nu^2>\over \rho^2} + {<\gamma^2>\over \Omega^2}) - \int_ {\tau '}^{\tau ''}{\vert N - \kappa \vert \over \sigma^2}d\tau \Bigr]\over \sqrt{sin\Bigl({N\over \sqrt{2}}\sqrt{1 + {i8\over 3\pi\Omega^2}}(\tau '' - \tau ')\Bigr)sin\Bigl(\sqrt{8}N\sqrt{1 - {i\over 6\pi\rho^2}}(\tau '' - \tau ')\Bigr)}}, \end{eqnarray} 
\noindent here $S = S_{\alpha} + S_{\beta}$
\bigskip

\section{Discussion and conclusions.}
\bigskip

In order to obtain in (25) a non-vanishing propagator several conditions must be fulfilled, one of them is that $\kappa(\tau)$ has to be almost a constant. Otherwise the term $\int_ {\tau '}^{\tau ''}{\vert N - \kappa \vert \over \sigma^2}d\tau$ generates an exponential decrease in the integrand of (25). From now on let us consider $\kappa(\tau) = \kappa = const$.

We proceed to define $t = \kappa (\tau'' - \tau')$ and $T = N(\tau'' - \tau')$, then (25) reduces to

{\setlength\arraycolsep{2pt}\begin{eqnarray} 
U_{[t, \nu, \gamma]}(q'', q') = \sqrt{{9\over 8}}\Bigl[(1 + {i8\over 3\pi\Omega^2})(1 - {i\over 6\pi\rho^2})\Bigr]^{1\over 4}\nonumber\\
\times \int{exp\Bigl[iS + T{3i\pi \over 4} -\vert T\vert({<\nu^2>\over \rho^2} + {<\gamma^2>\over \Omega^2}) - {\vert T - t \vert \over \sigma^2} \Bigr]\over \sqrt{sin\Bigl({T\over \sqrt{2}}\sqrt{1 + {i8\over 3\pi\Omega^2}}\Bigr)sin\Bigl(\sqrt{8}T\sqrt{1 - {i\over 6\pi\rho^2}}\Bigr)}}dT.
 \end{eqnarray}  

Clearly, $\kappa$ is an estimation of the lapse function $N$. Therefore, a gauge invariant physical time $t$ emerges as consequence of the measurement of the lapse function $N$ by higher mulipoles of matter, and from the form of the metric (1) we see that the physical time $t = (\tau'' - \tau')\kappa$ is indeed an estimation of the duration of the interval $[\tau', \tau'']$, 
while the error done in its measurement has the value $\Delta t = \sigma^2$.

We have constructed Halliwell's propagator for the case of a Mixmaster universe with small but non-vanishing anisotropy. Afterwards, in the context of the Decoherence Model, we have introduced in this system a self-measurement process, in which higher multipoles of matter act as environment for the superspace variables that in this proposal play the role of collective variables. 

Employing the Restricted Path Integral Formalism we have also calculated Halliwell's modified propagator, which appears as a consequence of this self-measurement process. 
This formalism has enabled us to take into account the influence of the measuring device without knowing the actual scheme of measurement. 

We have shown that a gauge invariant physical time appears as consequence of this self-measurement process. 

\Large{\bf Acknowledgments.}\normalsize
\bigskip

This work was partially supported by CONACYT Grant No. 3544--E9311. 
The author would also like to thank A. Mac{\'\i}as for the fruitful discussions on the subject.

\end{document}